\documentclass[
preprint, 
floatfix,aps,prd,showpacs,amsmath,amssymb,nofootinbib,preprintnumbers,superscriptaddress]{revtex4}

\usepackage{graphicx}

\newcommand{\simg}{\gtrsim}
\newcommand{\siml}{\lesssim}
\newcommand{\bib}{\bibitem}
\newcommand{\gravitino}{\psi_{3/2}}
\newcommand{\beq}{\begin{equation}}
\newcommand{\eeq}{\end{equation}}
\newcommand{\beqa}{\begin{eqnarray}}
\newcommand{\eeqa}{\end{eqnarray}}

\begin{document}

\begin{flushright}
ICRR-Report-575-2010-8\\
\end{flushright}

\title{ Prospects for Direct Detection of Inflationary Gravitational
  Waves by Next Generation Interferometric Detectors }

\author{Sachiko Kuroyanagi} \email[]{skuro@icrr.u-tokyo.ac.jp}
\affiliation{Institute for Cosmic Ray Research, University of Tokyo,
  Chiba 277-8582, Japan}

\author{Takeshi Chiba} \affiliation{Department of Physics, College of
  Humanities and Sciences, Nihon University, Tokyo 156-8550, Japan}

\author{Naoshi Sugiyama} \affiliation{Department of Physics and KMI, Nagoya
  University, Chikusa, Nagoya 464-8602, Japan} \affiliation{Institute
  for Physics and Mathematics of the Universe, University of Tokyo,
  Chiba 277-8582, Japan}

\begin{abstract}
  We study the potential impact of detecting the inflationary
  gravitational wave background by the future space-based
  gravitational wave detectors, such as DECIGO and BBO.  The
  signal-to-noise ratio of each experiment is calculated for
  chaotic/natural/hybrid inflation models by using the precise
  predictions of the gravitational wave spectrum based on numerical
  calculations.  We investigate the dependence of each inflation model
  on the reheating temperature which influences the amplitude and
  shape of the spectrum, and find that the gravitational waves could
  be detected for chaotic/natural inflation models with high reheating
  temperature.  From the detection of the gravitational waves, a lower
  bound on the reheating temperature could be obtained.  The
  implications of this lower bound on the reheating temperature for
  particle physics are also discussed.
\end{abstract}

\pacs{98.80.Cq, 04.30.-w}

\maketitle

\section{Introduction}
Gravitational wave, which is the gravitational counterpart of
electromagnetic wave, has eluded detection. Since gravitational waves
interact very weakly with matter, the Universe viewed with the
gravitational waves appears much transparent than that with photons.
Therefore, the detection of the gravitational waves gives us a
snapshot of the very early Universe.  Since we now already know that
the cosmic microwave background (CMB) spectrum has been a powerful
tool for cosmology and the analysis of it has revealed the various
components of the Universe, it is not difficult to imagine that
another new era of cosmology will come after the discovery of
gravitational waves.  At the time of the discovery of CMB in 1965, who
could have imagined the present situation of cosmology. It is
important to prepare for the coming age of another precision
cosmology.

Among the processes that occurred in the very early Universe,
inflation (including reheating) is the most important epoch.
Inflation was proposed as the most natural solution to the
difficulties of the standard big-bang cosmology, such as the horizon
problem and the flatness problem \cite{inflation}. It also generates
primordial gravitational waves (tensor modes) \cite{staro} as well as
the primordial density fluctuations (scalar modes) \cite{pert}. The
latter has already been observed as the cosmic microwave background
anisotropies by the Cosmic Background Explorer satellite
\cite{Bennett:1996ce} and by the Wilkinson Microwave Anisotropy Probe
(WMAP) satellite \cite{Bennett:2003bz}. The detection of the
gravitational waves generated during inflation could determine the
energy scale of inflation directly, which is essential to explore the
physics behind inflation \cite{Kuroyanagi:2009br}.

In Ref. \cite{kcs}, we calculated for the first time the most precise
power spectrum of the inflationary gravitational wave background for
several inflation models with perturbative reheating.  Our numerical
approach has enabled us to obtain the precise amplitude of the
spectrum over all frequencies.  We compute the spectrum all through
the evolution of the Hubble expansion rate, which determines the
amplitude of the gravitational waves, by following the dynamics of the
inflaton scalar field and its decay into radiation (reheating).  We
fully take into account of the changes of the effective number of
degrees of freedom during the radiation-dominated era which cause the
suppression of the spectrum at high frequencies \cite{Schwarz:1997gv}.
Therefore, all factors which affect the amplitude of the gravitational
wave background spectrum are precisely reflected in our calculation.

In this paper, we aim to forecast the future prospects of the direct
detection of the gravitational wave background by next generation
satellite detectors, such as the DECi-hertz Interferometer
Gravitational wave Observatory (DECIGO)
\cite{Seto:2001qf,Kawamura:2006up} and Big-Bang Observer (BBO)
\cite{bbo}.  While the measurement of CMB B-mode polarization, which
is the indirect signature of primordial gravitational waves, probes
gravitational waves at the present horizon scale, the direct detection
observes gravitational waves at a much smaller scale, which would
contain unique information about the very early Universe.  In
particular, we should stress that the shape of the gravitational wave
spectrum around the target frequency ($\sim 0.1$Hz) is sensitive to
the physics of reheating \cite{Nakayama:2008ip}.  In
Ref. \cite{Nakayama:2008wy}, the detectability of the gravitational
wave background is estimated taking into account the dependence of the
spectrum shape on reheating temperature.  Here, we reexamine the
detectability by calculating the signal-to-noise ratio (SNR) with the
precise prediction of the spectrum amplitude and the specific noise
spectra of DECIGO and BBO.  We study the dependence of the spectrum on
reheating temperature as well as on models of inflation.  In addition,
we discuss the implications of a possible lower bound of the reheating
temperature obtained from the future direct detection by DECIGO/BBO
for particle physics.

Note that our investigation is carried out assuming reheating via
perturbative decay of the inflaton field, namely we do not consider
nonperturbative effects during reheating, called preheating
\cite{preheatB,preheatF}.  If the nonperturbative effects are dominant
in the reheating process, the picture of reheating and its effect on
the gravitational wave background are significantly different from
those of perturbative reheating
\cite{preheatGW1,preheatGW2,preheatGW3}.  In this paper, since we
would like to provide a conservative estimate of the gravitational
wave background and its detectability, we only consider perturbative
processes which always exist and are directly related with the
reheating temperature.

The outline of this paper is as follows.  In Sec. \ref{GW}, the
signal-to-noise ratio expected in future experiments is estimated for
four inflation models; chaotic inflation with quadratic and quartic
potentials, natural inflation and hybrid inflation.  First of all, we
present the method to calculate the signal-to-noise ratio and the
current specification design of DECIGO and BBO.  Then the procedure of
the numerical calculation is described briefly, which is used to
obtain the amplitude of the spectrum.  In Sec. \ref{detect}, we
calculate the signal-to-noise ratio using the detailed experimental
specification and the precise prediction of the spectrum amplitude.
The implications of the lower limit of the reheating temperature on
particle physics are also discussed in relation with the gravitino
problem in Sec. \ref{particle}.  A summary is given in Sec. \ref{sum}.

\section{Estimation method of the signal-to-noise ratio}
\label{GW}

\subsection{Correlation analysis for detection of a stochastic
  gravitational wave background} 
Cosmological gravitational waves are described as tensor perturbations
in the Friedmann-Robertson-Walker metric as
$ds^2=-dt^2+a^2(t)(\delta_{ij}+h_{ij})dx^idx^j$, where $a(t)$ is the
scale factor of the Universe.  The tensor perturbation $h_{ij}$ can be
expanded into its Fourier components as
\begin{equation}
h_{ij}(t,\textbf{x})=\sum_{\lambda=+,\times}^{}\int\frac{d^3k}{(2\pi)^{3/2}}\epsilon_{ij}^{\lambda}
(\textbf{k})h_\textbf{k}^{\lambda}(t)e^{i\textbf{k}\cdot\textbf{x}},
\end{equation}
where the polarization tensors $\epsilon_{ij}^{+,\times}$ satisfy
symmetric and transverse-traceless conditions and are normalized as
$\sum_{i,j}^{}\epsilon_{ij}^{\lambda}(\epsilon_{ij}^{\lambda^{\prime}})^*=2\delta^{\lambda\lambda^{\prime}}$.
The intensity of a stochastic gravitational wave background is
characterized by the dimensionless quantity, $\Omega_{\rm
  GW}\equiv(d\rho_{\rm GW}/d\ln k)/\rho_c$, where the critical density
of the Universe is defined as $\rho_{c}\equiv3H^2/8\pi G$ with the
Hubble expansion rate, $H=(da/dt)/a$.  The energy density of the
gravitational waves $\rho_{\rm GW}$ is given from the 00-component of
the stress-energy tensor as $\rho_{\rm GW}=\langle (\partial_t
h_{ij})^2+(\vec{\nabla} h_{ij}/a)^2\rangle/(64\pi G)$.  Then
$\Omega_{\rm GW}$ can be expressed in terms of the Fourier component
$h_\textbf{k}^{\lambda}$ as \cite{Maggiore:1999vm}
\begin{equation}
\Omega_{\rm GW}=\frac{1}{12}\left(\frac{k}{aH}\right)^2\frac{k^3}{\pi^2}\sum_{\lambda}^{}|h_\textbf{k}^{\lambda}|^2.
\end{equation}

In future satellite missions like DECIGO and BBO, the analysis of a
stochastic gravitational wave background would be performed by taking
the cross correlation of the outputs of gravitational wave detectors
\cite{Kudoh:2005as,chiba,Seto:2005qy}.  The signal-to-noise ratio for
the correlation analysis is given in terms of an expected
(theoretical) form of $\Omega_{\rm GW}(f)$, and the functions related
to the experiment design, such as the noise spectrum $S_{I,J}(f)$ and
the overlap reduction function $\gamma_{IJ}(f)$ as \cite{Allen:1997ad}
\begin{equation}
[{\rm SNR}]^2=2\left(\frac{3H_0^2}{10\pi^2}\right)^2T_{\rm obs}\sum_{(I,J)}
\int^{\infty}_0df\frac{|\gamma_{IJ}(f)|^2\Omega_{\rm GW}^2(f)}{f^6S_I(f)S_J(f)},
\end{equation}
where $f=k/2\pi$ is the frequency of gravitational waves, $H_0$ is the
present Hubble expansion rate, $T_{\rm obs}$ is the duration of the
observation time.  The subscripts $I$ and $J$ refer to independent
signals obtained at each detector, or observables generated by
combining the detector signals.

DECIGO is planned to be a Fabry-Perot Michelson interferometer with an
arm length of $L=1.0\times 10^3$km \cite{Seto:2001qf,Kawamura:2006up}.
In this case, the SNR is calculated with the noise spectral density of
the two interferometers, which are assumed to be the same and given by
\cite{Kudoh:2005as,chiba}
\begin{equation}
S_1(f)=S_2(f)=S_{\rm shot}+S_{\rm accel}+S_{\rm rad},
\end{equation}
where the shot noise is given as $S_{\rm shot}=5.29\times
10^{-42}(1+f^2/f_c^2)(L/{\rm km})^{-2} {\rm Hz}^{-1}$, the
acceleration noise is $S_{\rm accel}=4.0\times 10^{-46}(f/{\rm
    Hz})^{-4}(L/{\rm km})^{-2}{\rm Hz}^{-1}$ and the radiation
  pressure noise is $S_{\rm rad}=3.6\times 10^{-51}(f/{\rm
    Hz})^{-4}(1+f^2/f_c^2)^{-1}{\rm Hz}^{-1}$. \footnote{A major
  improvement of the sensitivity is under consideration. }  The cutoff
frequency is given by $f_c=1/(4{\cal F}L)$ with the fineness for the
DECIGO detector, ${\cal F}=10$.

On the other hand, BBO would adopt a technique called time-delay
interferometry, in which new variables ($I=A,E,T$) are constructed to
cancel the laser frequency noise.  The noise transfer functions for
the time-delay interferometry variables are given as
\cite{Prince:2002hp,LISA}
\begin{eqnarray}
S_A(f)=S_E(f)=8\sin^2(\hat{f}/2)[(2+\cos\hat{f})S_{\rm shot}\nonumber\\
+2(3+2\cos\hat{f}+\cos(2\hat{f}))S_{\rm accel}],
\end{eqnarray}
\begin{equation}
S_T(f)=2[1+2\cos\hat{f}]^2[S_{\rm shot}+4\sin^2(\hat{f}/2)S_{\rm accel}],
\end{equation}
where $\hat{f}=2\pi Lf$.  In the case of BBO, the arm length is
$L=5.0\times 10^4$km, and the noise functions are $S_{\rm
  shot}=2.0\times 10^{-40}/(L/{\rm km})^2{\rm Hz}^{-1}$ and $S_{\rm
  accel}=9.0\times 10^{-40}/(2\pi f/{\rm Hz})^4/(2L/{\rm km})^2{\rm
  Hz}^{-1}$.

The overlap reduction function $\gamma_{IJ}(f)$ can be calculated by
taking into account of the relative locations and orientations of the
detectors.  We use the results of Ref. \cite{Kudoh:2005as} for
FP-DECIGO, and of Ref. \cite{Corbin:2005ny} for BBO.

\subsection{Numerical calculation for the spectrum of the gravitational wave background}
We briefly describe the method we used to obtain the theoretical
prediction for the amplitude of the spectrum $\Omega_{\rm GW}$ in our
previous work (For details, see Ref. \cite{kcs}).  We numerically
solve the evolution equation for gravitational waves, which is derived
from the perturbed Einstein equation under the
Friedmann-Robertson-Walker metric,
\begin{equation}
\ddot{h}_\textbf{k}^{\lambda}+3H\dot{h}_\textbf{k}^{\lambda}+\frac{k^2}{a^2}h_\textbf{k}^{\lambda}=0,
\label{heq2}
\end{equation}
where the over dot describes the time derivative.  The initial
condition is randomly taken from the Bunch-Davis vacuum, of which
variance is given as
\begin{equation}
|h_\textbf{k}^{\lambda}|^2=\frac{16\pi}{2ka^2m_{\rm Pl}^2},
\label{initial}
\end{equation}
where $m_{\rm Pl}=G^{-1/2}$ denotes the Planck mass.

The important point of our evaluation of the spectrum is that we
compute the evolution of the gravitational waves with following all
through the history of the cosmic expansion from inflation to the
present epoch.  It allows us to calculate the spectrum with no use of
the slow-roll approximation, which overestimates the amplitude of the
spectrum at the direct detection scale in some inflation models.  The
spectrum, which is Taylor-expanded around the CMB scale, can
overestimate the amplitude by $10-20\%$ \cite{kcs}.  The numerical
approach also enables us to precisely evaluate the effect of the
changes in the relativistic degrees of freedom during the
radiation-dominated era.

If we assume reheating is proceeded by perturbative decay of the
inflaton field into light fermions, all the processes from inflation
to the end of reheating can be calculated by simultaneously solving
the following equations,
\begin{eqnarray}
  \ddot{\phi}+(3H+\Gamma)\dot{\phi}+\frac{\partial V(\phi)}{\partial\phi}=0,\label{reheat1}\\
  \dot{\rho}_r+4H\rho_r=\Gamma\rho_{\phi},\label{reheat2}\\
  H^2=\frac{8\pi}{3m_{\rm Pl}^2}(\rho_{\phi}+\rho_r),\label{reheat3}
\end{eqnarray}
where $\Gamma$ is the decay rate of the scalar field into radiation,
$V(\phi)$ is the potential of the scalar field, $\rho_r$ is the energy
density of the radiation, and the energy density of the scalar field
is given as $\rho_{\phi}=\dot{\phi}^2/2+V(\phi)$. 

During inflation, the Hubble expansion rate is determined by the
dynamics of the scalar field $\phi$ which drives inflation.  The
Universe enters a reheating phase after inflation, and the scalar
field oscillates at the bottom of the potential decaying into
radiation.  During this phase, the Universe evolves like a
matter-dominated Universe in most of the inflation models, and turns
into a radiation-dominated era after it ends.  The exception is, for
example, the case where the potential has a $\lambda\phi^4$ shape at
its bottom.  In this case, the Universe behaves as a
radiation-dominated Universe and connects to the subsequent
radiation-dominated era with no change in the Hubble expansion rate.

If the matter-dominated reheating phase exists before the
radiation-dominated era, the inflationary gravitational wave spectrum
is suppressed at high frequencies \cite{Nakayama:2008ip,
  Nakayama:2008wy}.  The suppression is seen on the modes which enter
the horizon during reheating, since the matter-dominated phase induces
frequency dependence of $f^{-2}$ on the spectrum while a
radiation-dominated era gives a flat spectrum $\propto f^0$.  The
characteristic frequency, where the change of the frequency dependence
from $f^{-2}$ to $f^0$ arises, is given in terms of the temperature of
the Universe at the end of reheating as \cite{Kamionkowski:1993fg}
\begin{equation}
  f_{\rm RH}\simeq 0.3\left(\frac{T_{\rm RH}}{10^7{\rm GeV}}\right)
  \left(\frac{g_{*,{\rm RH}}}{220}\right)^{1/2}\left(\frac{g_{*s,{\rm RH}}}{220}\right)^{-1/3} {\rm Hz},
\label{freq}
\end{equation}
where $g_*$ and $g_{*s}$ are the effective number of relativistic
degrees of freedom contributing to the radiation density and the
entropy density. The subscript "RH" denotes the value at the end of
reheating.  We take $g_{*,{\rm RH}}$ to be $\sim 220$, which includes
degrees of particles in minimal supersymmetric standard model.  Since
both $g_*$ and $g_{*s}$ take the same value before electron-positron
annihilation, $\sim 0.1$MeV, where the temperature of neutrinos
becomes lower than that of photons, $g_{*s,{\rm RH}}$ is also taken to
be $220$.  In the case of the perturbative decay, the reheating
temperature $T_{\rm RH}$ is related to the decay rate $\Gamma$ as
\cite{Kolb}
\begin{equation}
T_{\rm RH}\simeq
g_{*,{\rm RH}}^{-(1/4)}\left(\frac{45}{8\pi^3}\right)^{1/4}(m_{\rm Pl}\Gamma)^{1/2}.
\label{TRH}
\end{equation}

Note that the above expressions are valid only for the case where the
mass of the interacting fermion is much smaller than the mass of the
inflaton field and also the coupling constant is sufficiently small.
Otherwise, parametric resonance between the two interacting fields may
give nonperturbative creation of particles and dramatically shorten
the time scale of reheating.  We do not deal with such
nonperturbative particle production during reheating, called
preheating \cite{preheatB}.  If the nonperturbative growth is
sufficient to reheat the Universe, the $f^{-2}$ dependence would not
arise on the inflationary gravitational wave background and it may
rather be important to look at gravitational waves originating from
large inhomogeneities in the matter field during the nonperturbative
stage \cite{preheatGW1,preheatGW2,preheatGW3}.  Such gravitational
waves typically have very high frequency beyond the sensitivity
bandwidth of DECIGO and BBO, but some models predict an infra-red tail
which overlaps or exceeds the inflationary gravitational wave
background at detectable frequency.  However, in this paper, our focus
is on the effect of reheating on the inflationary gravitational wave
background and consider only the case of reheating with perturbative
fermionic decay.  For reheating via fermionic decay, the rapid
particle production is suppressed by Pauli blocking, while for
reheating via bosonic decay could be easily completed by the
exponential growth of the number of particles.  Although the effect of
the nonperturbative decay can still be important for the fermionic
case \cite{preheatF} and could change the time scale of reheating, our
investigation, after all, turns out to be only the weak coupling case,
where the nonperturbative effects are negligibly small.  The detailed
analysis of the effects of nonperturbative processes like preheating
on the inflationary gravitational wave background will be left to a
future work.

After reheating ends, the Universe enters a radiation-dominated phase.
Including the effects of $g_*(T)$ and $g_{*s}(T)$, the Hubble
expansion rate can be written as \cite{Watanabe:2006qe}
\begin{equation}
H^2=H_0^2\left[\left(\frac{g_*(T)}{g_{*,0}}\right)\left(\frac{g_{*s}(T)}{g_{*s,0}}\right)^{-4/3}\Omega_r a^{-4}+\Omega_m a^{-3}+\Omega_{\Lambda}\right],
\label{Hubble_g}
\end{equation}
where $0$ denotes values at the present time, which are given by
summing the contributions from photons and neutrinos: $g_{*,0}=3.36$
and $g_{*s,0}=3.90$.  The change of the Hubble expansion rate affects
the evolution of the inflationary gravitational waves and causes
damping at the frequency where the direct detection experiments are
targeting.  The suppression is about $(g_*(T=10^7{\rm
  GeV})/g_{*,0})(g_{*s}(T=10^7{\rm
  GeV})/g_{*s,0})^{-4/3}=(220/3.36)(220/3.90)^{-4/3}\sim 0.3$.

We calculate the spectrum of the gravitational waves by numerically
solving the evolution of each mode with Eq. (\ref{heq2}).  At the same
time, we follow the evolution of the Hubble expansion, which is
calculated by Eqs. (\ref{reheat1}), (\ref{reheat2}) and
(\ref{reheat3}) during the inflation and reheating phase, and by
Eq. (\ref{Hubble_g}) after reheating.  In this paper, the energy
density of radiation is taken to be $\Omega_r h^2=4.15\times 10^{-5}$
and the other cosmological parameters are given by the maximum
likelihood values from the combined constraints of the WMAP 7 yr,
BAO, and supernova data\cite{Komatsu:2010fb}: matter density $\Omega_m
h^2=0.1344$, cosmological constant density $\Omega_\Lambda=0.728$,
amplitude of curvature perturbations $\Delta_{\cal R}^2=2.45\times
10^{-9}$, and the Hubble parameter $h=0.702$.

\section{Predictions for the detectability in future experiments}
\label{detect}

\subsection{Signal-to-Noise Ratio}

\begin{figure}
 \begin{center}
  \includegraphics[width=0.48\textwidth]{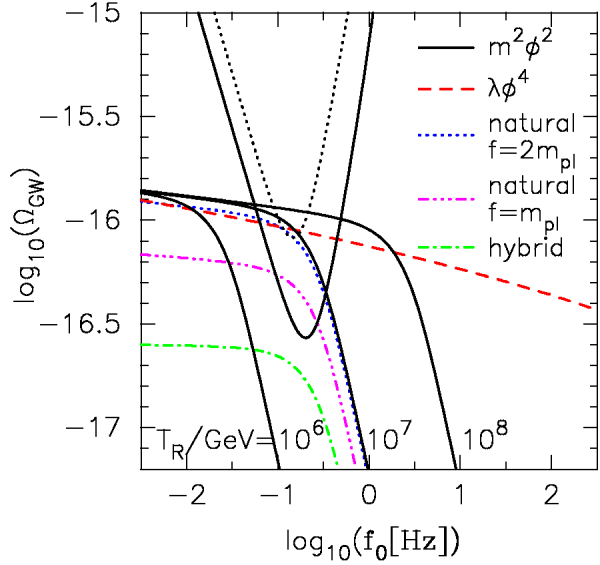}
  \caption{\label{spectrum} Spectra of the gravitational wave
    background for different inflation models, shown with the
    sensitivity curves of DECIGO (dotted) and BBO (solid).  The
    spectra are calculated assuming $T_{\rm RH}=10^7$GeV.  The cases
    of $T_{\rm RH}=10^6$GeV and $10^8$GeV are also plotted assuming
    the quadratic potential model.  Note that the spectrum lines mean
    the time-averaged value of $\Omega_{\rm GW}$. }
 \end{center}
\end{figure}

Now we forecast the detectability of the inflationary gravitational
wave background with DECIGO/BBO for several inflation models: chaotic
inflation with a quadratic and quartic potential, natural inflation
\cite{Freese:1990rb,Adams:1992bn}, and hybrid inflation
\cite{Linde:1993cn}. The potentials are respectively given by,
\begin{eqnarray}
{\rm quadratic:}&&V=\frac{1}{2}m^2\phi^2,\\
{\rm quartic:}&&V=\frac{1}{4}\lambda\phi^4,\\
{\rm natural:}&&V=\Lambda^4\left[1\pm\cos\left(\frac{N\phi}{f}\right)\right],\\
{\rm hybrid:}&&V=\frac{1}{4\lambda}(M^2-\lambda\sigma^2)^2
+\frac{1}{2}m^2\phi^2+\frac{1}{2}g^2\phi^2\sigma^2.
\end{eqnarray}
The spectra of these four models and the sensitivity curves of the
DECIGO and BBO experiments are shown in Fig. \ref{spectrum}.  The
normalization of the scalar perturbations $\Delta_{\cal
  R}^2=2.45\times 10^{-9}$ fixes the value of the potential
parameters.  It gives $m=1.72 \times 10^{13} $GeV for the quadratic
potential, $\lambda =1.54\times 10^{-13}$GeV for the quartic
potential, $\Lambda=2.04\times 10^{16}$GeV for the natural inflation
model with $N=1$ and $f=2m_{\rm Pl}$, $\Lambda=1.33\times 10^{16}$GeV
for the natural inflation model with $N=1$ and $f=m_{\rm Pl}$, and
$M=1.45\times 10^{16}$GeV for the hybrid inflation model with
$\lambda=1$, $g=8\times 10^{-4}$, and $m=2.5\times 10^{-7}m_{\rm Pl}$.
\footnote{Although there is much freedom in the choice of the
  parameters for hybrid inflation, here we take one example where the
  parameter $M$ corresponds to grand unification scale.}  The
parameter values are searched numerically by adjusting the resulting
Hubble expansion rate to be $H_0=100h$km/s/Mpc.  Note that these
parameters also slightly depend on the reheating temperature, since it
is related to the length of inflation.  The above values are obtained
when $T_{\rm RH}=10^7$GeV.

\begin{figure}
 \begin{center}
  \includegraphics[width=0.48\textwidth]{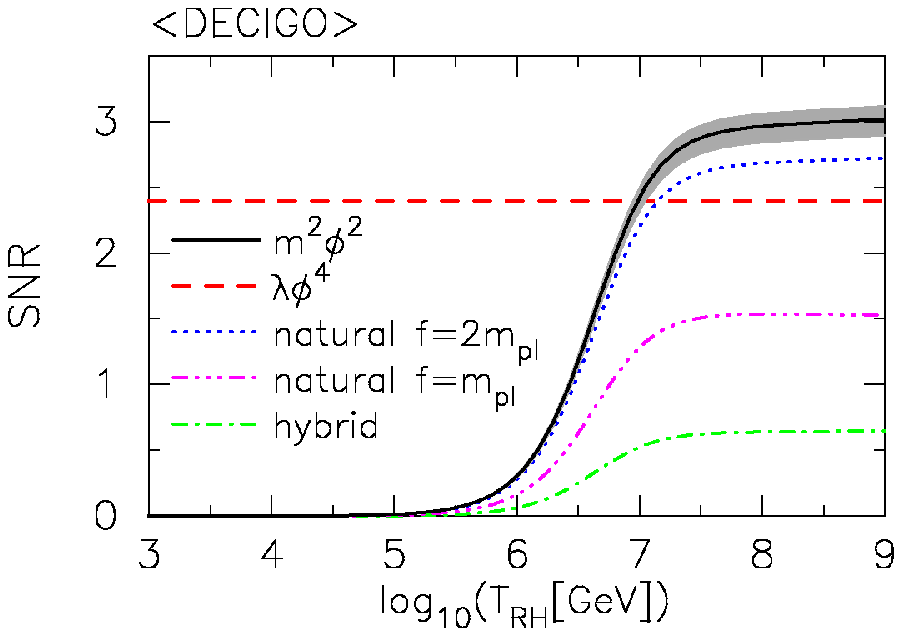}\\
  \includegraphics[width=0.48\textwidth]{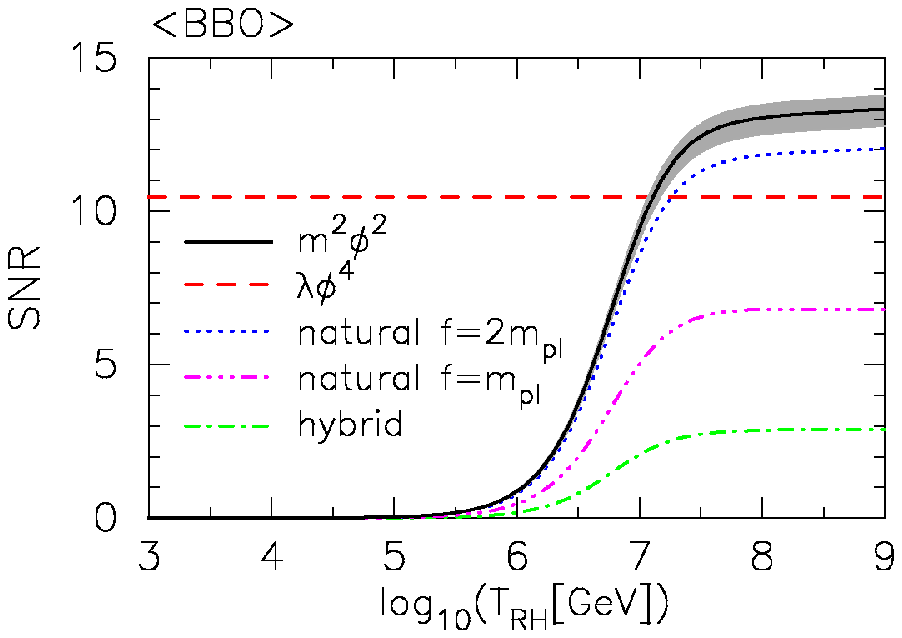}
  \caption{\label{SNR1} Signal-to-noise ratio vs reheating temperature
    calculated for the four different inflation models.  The upper
    panel shows the SNR for DECIGO and the lower panel is for BBO.
    The gray region shows the 1$\sigma$ uncertainty in the
    normalization $\Delta_{\cal R}^2=(2.441^{+0.088}_{-0.092})\times
    10^{-9}$ from WMAP 7 yr, which is used to determine the energy
    scale of inflation.}
 \end{center}
\end{figure}

In Fig. \ref{SNR1}, we show the SNR for cross-correlation analysis
expected with a 10-year observation by DECIGO and BBO.  The SNR
decreases significantly when the reheating temperature $T_{\rm RH}$ is
lower than $10^7$GeV, since the suppression due to the presence of the
reheating phase reduces the amplitude of the spectrum at the target
frequency of the direct detection experiments.  The only exception is
the case of the quartic potential, in which the Hubble expansion rate
behaves as a radiation-dominated Universe during the reheating phase
and the suppression does not arise.

In Table \ref{table}, we present the relation between the value of
$\Omega_{\rm GW}$ at the detection frequency, $f=0.2$Hz, and the
tensor-to-scalar ratio, which is usually evaluated at the CMB scale
{$k_{\rm CMB}=0.002{\rm Mpc}$} as
\begin{equation}
  r\equiv\frac{\Delta_{h}^2(k_{\rm CMB})}{\Delta_{\cal R}^2(k_{\rm CMB})}\simeq 16\epsilon,
\label{tsratio} 
\end{equation}
where the slow-roll parameter is defined as $\epsilon\equiv m_{\rm
  Pl}^2/(16\pi)(V^{\prime}/V)^2|_{k_{\rm CMB}=aH}$.
The reheating temperature
is set as $T_{\rm RH}=10^{9}$GeV which is so high that the suppression
does not arise at the detection frequency. As is clear from the
comparison between $m^2\phi^2$ and $\lambda\phi^4$ model, the
amplitude of the gravitational wave at the direct detection scale is
not proportional to the tensor-to-scalar ratio $r$ because the tilt of
the spectrum $n_T\simeq -2\epsilon$ and the higher order terms of the
Taylor-expansion becomes important when connecting the two different
scales \cite{Friedman:2006zt}.  This is prominent in models which
predict larger $r$.

For DECIGO, the inflationary gravitational background could be
detected with SNR$\geq 3$ if the inflation model is chaotic inflation
with $T_{\rm RH}\simg 10^7$GeV.  For BBO, the inflationary
gravitational background could be detected with SNR$\geq 5$ if the
inflation model is chaotic inflation with $T_{\rm RH}\simg 2\times
10^6$GeV or natural inflation with $f\simg m_{\rm Pl}$ and $T_{\rm
  RH}\simg 10^7$GeV.  Therefore, from the contraposition, if DECIGO
does not detect the inflationary gravitational wave background, the
chaotic inflation model will be excluded unless the reheating
temperature is lower than $10^7$GeV.  The same argument holds for BBO
except that it can apply to natural inflation.

\begin{table}
\begin{center}
\begin{tabular*}{0.9\textwidth}{@{\extracolsep{\fill}}lcccc}
  \hline
  \hline
  \rule[-3pt]{0pt}{14pt}Model & $r$ & $\Omega_{\rm GW}$ & SNR (BBO) & SNR (DECIGO)\\
  \hline
  \rule[0pt]{0pt}{12pt}$m^2\phi^2$ & 0.144 & $1.12\times 10^{-16}$ & 13.3 & 3.02\\
  $\lambda\phi^4$ & 0.262 & $8.75\times 10^{-17}$ & 10.5 & 2.40\\
  Natural ($f=2m_{\rm Pl}$) & 0.108 & $1.01\times 10^{-16}$ & 12.0 & 2.72\\
  Natural ($f=m_{\rm Pl}$) & 0.0406 & $5.70\times 10^{-17}$ & 6.79 & 1.53\\
  Hybrid & 0.0104 & $2.44\times 10^{-17}$ & 2.90 & 0.646\\[4pt]
  \hline
  \hline
\end{tabular*}
\caption{The tensor-to-scalar ratio $r$, the amplitude of the
  gravitational wave $\Omega_{\rm GW}$ at $f=0.2$Hz, and the
  signal-to-noise ratio in DECIGO and BBO for each inflation model.
  The reheating temperature is set as $T_{\rm
    RH}=10^{9}$GeV. \label{table}}
\end{center}
\end{table}

\begin{figure}
 \begin{center}
  \includegraphics[width=0.48\textwidth]{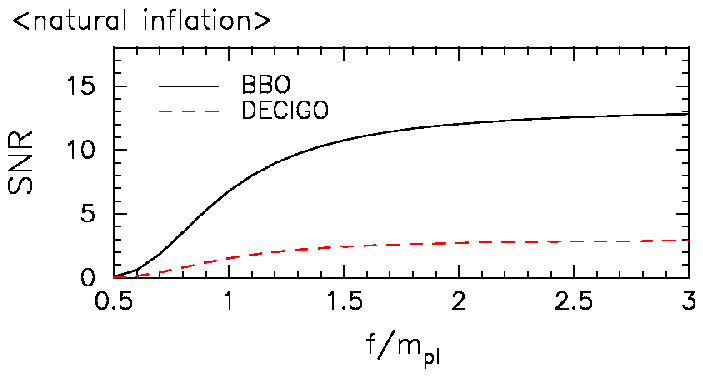}
  \caption{ \label{paraN} Parameter dependence of the signal-to-noise
    ratio for the natural inflation model.}
\bigskip
  \includegraphics[width=0.48\textwidth]{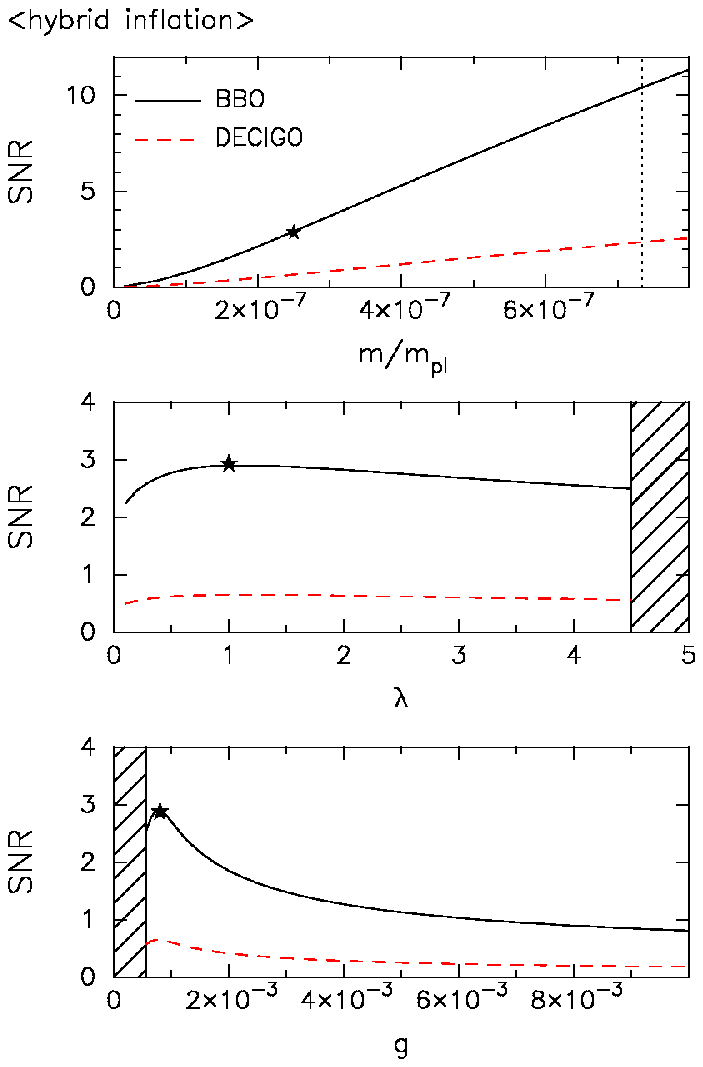}
  \caption{ \label{paraH} Parameter dependencies of the
    signal-to-noise ratio for the hybrid inflation model. The stars
    show the fiducial point, at which parameters are set as
    $\lambda=1$, $g=8\times 10^{-4}$, and $m=2.5\times 10^{-7}m_{\rm
      Pl}$. The dotted line in the top panel shows $y=1$.  The shaded
    region means $\sigma\neq0$ when the observable scale (the present
    Hubble horizon) exits the horizon.}
 \end{center}
\end{figure}

\subsection{Parameter Dependence}
We also investigate the parameter dependence of the SNR, shown in
Figs. \ref{paraN} and \ref{paraH} by setting the reheating temperature
to be $T_{\rm RH}=10^{9}$GeV.  For the natural inflation model, the
amplitude of the gravitational waves becomes larger as $f$ increase.
This can be interpreted as follows: Since we use the normalization of
the scalar perturbations $\Delta_{\cal R}^2=2.45\times 10^{-9}$, the
value of $\epsilon$ determines the amplitude of the gravitational
waves at the CMB scale, as we know
$\Delta_{h}^2=16\epsilon\Delta_{\cal R}^2$ from Eq. (\ref{tsratio}).
\footnote{Although this relation does not hold at scales well below
  the CMB scale, the amplitude at the direct detection scale is a
  monotonically increasing function of $\epsilon$ for the natural and
  hybrid inflation models, which predict relatively small slow-roll
  parameters.}

In the case of the natural inflation model with $N=1$, $\epsilon$ is
given as \cite{Savage:2006tr}
\begin{equation}
\epsilon=\frac{1}{16\pi}\left(\frac{m_{\rm Pl}}{f}\right)^2
\left[\frac{\sin(x)}{1+\cos(x)}\right]^2,
\label{epsilonN}
\end{equation}
where we define $x\equiv\phi/f$.  The field value can be written in
terms of the e-folding number, ${\cal N}\equiv\ln(a_{\rm end}/a)$, as
\cite{Freese:2004un}
\begin{equation}
\sin\left(\frac{x}{2}\right)=\sin\left(\frac{x_{\rm end}}{2}\right)
\exp\left(-\frac{m_{\rm Pl}^2}{16\pi f^2}{\cal N}\right).
\label{fieldN}
\end{equation}
The end of inflation $x_{\rm end}$ is defined at the point
where the slow-roll condition is violated, $\epsilon=1$, which gives
\begin{equation}
\cos(x_{\rm end})=\frac{1-16\pi(f/m_{\rm Pl})^2}{1+16\pi(f/m_{\rm Pl})^2}.
\label{endN}
\end{equation}
Combining Eqs. (\ref{epsilonN}) (\ref{fieldN}) and (\ref{endN}), we
find that $\epsilon$ is written as $\epsilon=1/[(1+\chi)\exp(2{\cal
  N}/\chi)-\chi]$ where $\chi\equiv16\pi f^2/m_{\rm Pl}^2$. This is an
increase function of $\chi$, which implies that $\Omega_{\rm GW}$
increases as $f$ increases.

In the standard picture of hybrid inflation that the $\phi$ field
determines the evolution of the Universe during inflation, the
potential can be recast as $V=\Lambda[1+(\phi/\mu)^2]$ for the
inflation stage, where we define $\Lambda\equiv M^2/(4\lambda)$ and
$\mu\equiv M^2/(m \sqrt{2\lambda})$ \cite{Smith:2005mm}.  Then the
slow-roll parameter is given as
\begin{equation}
\epsilon=\frac{1}{4\pi}\left(\frac{m_{\rm Pl}}{\mu}\right)^2\frac{y^2}{[1+y^2]^2}.
\label{epsilonH}
\end{equation}
The field value is given in terms of the e-folding number
\cite{Martin:2006rs},
\begin{equation}
y=\sqrt{W_0\left\{y_{\rm end}^2
    \exp\left[y_{\rm end}^2
      +\frac{\cal N}{2\pi}\left(\frac{m_{\rm Pl}}{\mu}\right)^2\right]\right\}},
\label{fieldH}
\end{equation}
where we define $y=\phi/\mu$ and $W_0(x)$ is the principal branch of
the Lambert function, which satisfies $x=W_0(x)e^{W_0(x)}$.  In the
case of hybrid inflation, inflation ends with the waterfall field
$\sigma$ rolling down when $\phi_{\rm end}=M/g$, which gives
\begin{equation}
y_{\rm end}=\frac{m\sqrt{2\lambda}}{Mg}.
\label{endH}
\end{equation}

The potential looks more like quadratic with increasing $m$, which is
the mass of the $\phi$ field.  When $y>1$, the inflation dynamics
becomes the same as the case of the quadratic potential.  So, we
focus our interest on the case of $y<1$.  The value of $m$ affects
both $\mu$ and $y$, which respectively decreases and increases with
increasing $m$.  From Eqs. (\ref{epsilonH}) (\ref{fieldH}) and
(\ref{endH}), we find $\epsilon$ becomes larger with decreasing $\mu$
and increasing $y$ for $0<y<1$.  Hence $\Omega_{\rm GW}$ increases as
$m$ increases as seen in the top panel of Fig. \ref{paraH}.

The parameter $\lambda$ determines the potential minimum of the
$\sigma$ field, $\sigma_{\rm min}=M/\sqrt{\lambda}$.  Since
observables are determined by the dynamics of the $\phi$ field in the
standard hybrid inflation, the value of $\lambda$ has little
effect on the amplitude of the gravitational waves as shown in the
middle panel of Fig. \ref{paraH}.

The parameter $g$ determines the value of $\phi_{\rm end}=M/g$, where
the waterfall field $\sigma$ starts to roll down.  The value of $y$ is
affected through $y_{\rm end}=\phi_{\rm end}/\mu$ which becomes
smaller as $g$ increases.  Thus, the slow-roll parameter $\epsilon$,
which is an increase function of $y$ for $0<y<1$, becomes smaller with
increasing $g$.  Hence, SNR $\propto\Omega_{\rm GW}$
decreases as $g$ becomes larger in Fig. \ref{paraH}.

For too large $\lambda$ or small $g$, the condition $m^2/\phi_{\rm
  end}^2=m^2M^2/g^2\ll M^4/\lambda$ \cite{Linde:1993cn} is not
satisfied, which results in that the waterfall field $\sigma$ starts to
roll down slowly before inflation is dominated by the vacuum energy of
the $\phi$ field.  The shaded region in the figure means $\sigma\neq0$
when the observable scale (the present Hubble horizon) exits the
horizon, which is not the standard behavior of the hybrid inflation
model.

\section{Implications of the Lower Limit of the Reheating Temperature for Particle Physics}
\label{particle}

In previous sections, we found that if inflation is the chaotic type
and its reheating temperature is higher than $10^7$ GeV, then the
inflationary stochastic gravitational wave background would be
detected by future space-based interferometric detectors like DECIGO
or BBO.  Unfortunately we have a small chance to determine the exact
value of the reheating temperature from the detection.  The detection
only allows us to set a lower bound on the reheating temperature.
However, this lower bound $10^7$ GeV provides useful and unique
information of the very early Universe.  In this section, as an
example, we consider the implications of the lower bound of the
reheating temperature for the nature of gravitino production in the
early Universe.

Many models of supersymmetry breaking, in the context of either
supergravity or superstring theory, predict the presence of scalar
fields with Planck-suppressed couplings and masses around or heavier
than the weak scale. These fields are generically called moduli. The
coherent oscillation of the modulus soon dominates the Universe, and
the late decay of the modulus results in very low reheating
temperature, upsetting the success of the big-bang nucleosynthesis
(BBN) [called the moduli problem \cite{moduli}] unless the modulus is
ultraheavy: $M_X\simg 100$ TeV.

Gravitino is the fermionic superpartner of graviton and has
Planck-suppressed interaction.  The gravitino, once produced, decays
with a very long lifetime if it is unstable.  The gravitino may cause
several problems in cosmology. For example, if the gravitino is
unstable ({if its mass is} $M_{3/2}\sim 100{\rm GeV}-100{\rm TeV}$),
the decay products of the gravitino would destroy the primordial light
elements by photodissociation and hadrodissociation, and thus spoil
the success of BBN (called the gravitino problem \cite{gravitino}).
Hence, the yield of the gravitinos $Y_{3/2}=n_{3/2}/s$ ($s$ is the
entropy density) should be constrained: $Y_{3/2}<Y_{3/2}^{\rm BBN}$,
where the value of $Y_{3/2}^{\rm BBN}$ depends on the gravitino mass.
According to the recent analysis \cite{kkm,kmy}, $Y_{3/2}^{\rm BBN}
\sim 10^{-16}$ for $M_{3/2}\sim 1$TeV and $Y_{3/2}^{\rm BBN} \sim
10^{-15}-10^{-13}$ for $M_{3/2}\sim 10-100$ TeV.  On the other hand,
if the gravitino is stable (for $M_{3/2}\siml 1$GeV), the gravitinos
can be cold/warm dark matter.  Hence the abundance of the gravitinos
$\Omega_{3/2}h^2$ should also be bounded in order not to exceed the
dark matter abundance $\Omega_{dm}h^2\simeq 0.134$ which is precisely
determined by the WMAP \cite{Komatsu:2010fb}.

Recently, it has been found that gravitinos are produced not only by
the thermal scatterings at the reheating \cite{buch} but also by the
decay of heavy scalar fields (for example, inflaton and moduli)
\cite{fumi}. Therefore, the "ultraheavy moduli solution" to the moduli
problem may cause instead a new gravitino problem by the moduli decay.

In the following, we consider the implication of the possible lower
bound of the reheating temperature on gravitino cosmology in the
context of chaotic/natural inflation. Then the nonthermal production
of the gravitinos is only due to the moduli decay \cite{asaka} since
the vacuum expectation value of the inflaton is vanishing for
chaotic/natural inflation (with $Z_2$ symmetry) \cite{kawasaki}.
A similar consideration (without moduli decay) is given in
\cite{Nakayama:2008ip}.

\subsection{Unstable Gravitino}

Firstly we consider the case of unstable gravitinos. The yield of the gravitinos produced by the thermal 
scatterings at the reheating temperature $T_{\rm RH}$ is estimated by \cite{buch,kkm}\footnote{The thermal 
production of gravitinos at $T_X$ is a factor of $\rho_X/\rho_{\phi}$ smaller than that at $T_{\rm RH}$ and hence may 
be negligible.}
\beqa
Y_{3/2}^{\rm TH}\simeq 1.4\times 10^{-12}\left(\frac{T_{\rm RH}}{10^{10}{\rm GeV}}\right).
\label{unstable:th}
\eeqa
Thus if the reheating temperature is found to be high ($T_{\rm RH}\simg 10^7$ GeV), 
it would immediately imply that large gravitino mass $M_{3/2}\simg 10$ TeV is favored \cite{kmy}. 

The yield of the gravitinos by the $X$ decay is given by \cite{fumi,asaka}
\beqa
Y_{3/2}^X\simeq \frac{1}{192\pi}\sqrt{\frac{90}{\pi^2g_*(T_X)}}\frac{d_{3/2}^2M_X^2}{T_X M_{\rm Pl}},
\label{unstable:x1}
\eeqa
where $M_{\rm Pl}=m_{\rm Pl}/\sqrt{8\pi}$ is the reduced Planck mass 
and $d_{3/2}\simeq |\langle X\rangle |/M_{\rm Pl}$ is related to the partial decay rate of the process 
$X\rightarrow \gravitino +\gravitino$ as
\beqa
\Gamma_{3/2}=\frac{d_{3/2}^2}{288\pi}\frac{M_X^3}{M_{\rm Pl}^2}.
\eeqa
$T_X$ is the reheating temperature by $X$ decay with the decay rate 
$\Gamma_X\simeq M_X^3/(8\pi M_{\rm Pl}^2)$, \footnote{We hereby have fixed the order one coefficient. } and is given by
\beqa
T_X=\left(\frac{90}{\pi^2g_*(T_X)}\right)^{1/4}\sqrt{\Gamma_X M_{\rm Pl}}
=5.8\times 10^4{\rm GeV} \left(\frac{M_X}{10^{10}{\rm GeV}}\right)^{3/2},
\eeqa
where $g_*(T_X)$ is the effective relativistic degrees of freedom at $T=T_X$ and we use
$g_*(T\simg 1{\rm TeV})\simeq 220$. 
Then Eq. (\ref{unstable:x1}) becomes 
\beqa
Y_{3/2}^X\simeq 2.4 \times 10^{-7}\left(\frac{M_X}{10^{10}{\rm GeV}}\right)^{1/2}d_{3/2}^2
\label{unstable:x2}
\eeqa

Therefore, from the success of the BBN ($Y_{3/2}^X<Y_{3/2}^{\rm BBN}$), 
the upper bound on the moduli mass is found: \footnote{
Here we have neglected the effect of dilution by possible entropy productions by moduli decay. 
The effect would decrease the thermal yield by the dilution factor $F^{-1}$ and 
would weaken the bound on $M_X$ by $F^2$ \cite{Nakayama:2008ip}. }
\beqa
M_X\siml 2\times 10^{-3} {\rm GeV} 
d_{3/2}^{-4}\left(\frac{Y_{3/2}^{\rm BBN}}{10^{-13}}\right)^2.
\eeqa

\subsection{Stable Gravitino}

Next, we consider the case of stable gravitinos. The present-day abundance of 
the gravitinos produced by the thermal scatterings at the reheating is given by \cite{buch}
\beqa
\Omega_{3/2}^{TH}h^2\simeq 0.27\left(\frac{T_{\rm RH}}{10^{8}{\rm GeV}}\right)
\left(\frac{M_{3/2}}{1{\rm GeV}}\right)^{-1},
\label{stable:th}
\eeqa
where we have set the gluino mass $M_{\widetilde{g}}=1$TeV. 
{}From Eq. (\ref{unstable:x2}), the abundance of the gravitinos produced by the $X$ decay is given by
\beqa
\Omega_{3/2}^Xh^2=\frac{M_{3/2}Y_{3/2}^X}{\rho_{cr}/s_0}h^2\simeq 
6.8\times 10^{1}d_{3/2}^2 \left(\frac{M_{X}}{10^{10}{\rm GeV}}\right)^{1/2}
\left(\frac{M_{3/2}}{1{\rm GeV}}\right),
\label{stable:x}
\eeqa
where $\rho_{cr}$ is the present critical density of the Universe and $s_0$ is the present 
entropy density. 
The total abundance should satisfy the bound 
$\Omega_{3/2}h^2=\Omega_{3/2}^{\rm TH}h^2+\Omega_{3/2}^Xh^2\leq \Omega_{dm}h^2\simeq 0.134$. 

Therefore, if $T_{\rm RH}>T_{\rm RH}^{\rm gw}$ from gravitational wave experiments, then, 
using Eq. (\ref{stable:th}) from $\Omega_{3/2}^{\rm TH}h^2<\Omega_{dm}h^2$, we find the {\it lower bound} on the 
gravitino mass: 
\beqa
M_{3/2}> 0.21 {\rm GeV}\left(\frac{T_{\rm RH}^{\rm gw}}{10^{7}{\rm GeV}}\right)
\left(\frac{\Omega_{dm}h^2}{0.134}\right)^{-1}.
\eeqa
Moreover, from the geometric mean, 
\beqa\Omega_{dm}h^2\geq\Omega_{3/2}h^2=\Omega_{3/2}^{\rm TH}h^2+\Omega_{3/2}^{X}h^2\geq 
2\sqrt{\Omega_{3/2}^{\rm TH}\Omega_{3/2}^X}h^2\simeq 
8.6 \left(\frac{T_{\rm RH}}{10^{8}{\rm GeV}}\right)^{1/2} 
\left(\frac{M_X}{10^{10}{\rm GeV}}\right)^{1/4}d_{3/2},
\eeqa
we obtain the {\it upper bound} on the moduli mass: 
\beqa
M_X< 6\times 10^{4}{\rm GeV} \left(\frac{T_{\rm RH}^{\rm gw}}{10^{7}{\rm GeV}}\right)^{-2}
\left(\frac{\Omega_{dm}h^2}{0.134}\right)^{4}d_{3/2}^{-4}.
\eeqa

\section{Summary}
\label{sum}
The direct detection of the inflationary gravitational wave background
by the next generation space-based gravitational wave missions may
enable us to explore the early Universe more deeply than current
observations.  In this paper, the detectability of the inflationary
gravitational wave background in the future experiments, DECIGO and
BBO, has been estimated using our precise predictions for the
spectrum.  We have considered several inflation models and have taken
into account the effect of the reheating temperature which determines
the frequency where the signature of reheating arises on the
gravitational wave spectrum.

We have found that DECIGO could detect the inflationary gravitational
background with SNR$\geq 3$, for the chaotic inflation model with
$T_{\rm RH}\simg 10^7$GeV.  The higher sensitivity of BBO makes
possible a detection with SNR$\geq 5$, for the chaotic inflation model
with $T_{\rm RH}\simg 2\times 10^6$GeV or the natural inflation model
with $f\simg m_{\rm Pl}$ and $T_{\rm RH}\simg 10^7$GeV.  This means,
conversely, that non detection of the inflationary gravitational wave
background by DECIGO would exclude the chaotic inflation model unless
the reheating temperature is lower than $10^7$GeV.  BBO would exclude
it unless $T_{\rm RH}\siml 2\times 10^6$GeV and further exclude the
natural inflation model with $f\simg m_{\rm Pl}$ unless $T_{\rm
  RH}\siml 10^7$GeV.

We have also discussed the implications of the possible lower bound of
the reheating temperature on gravitino cosmology in the context of
chaotic/natural inflation.  Taking into account of both thermal and
nonthermal production of gravitinos, we find that from the lower
bound on the reheating temperature we could obtain a lower bound on
the gravitino mass and an upper bound on the moduli mass.  These bounds
may provide information regarding the gravitino mass and the moduli
mass complementary to collider experiments. Thus, in future, the
direct detection of the inflationary gravitational wave background
could play a crucial role in probing not only the history of the early
Universe but also particle physics.

\section*{Acknowledgments}
The authors are grateful to N. Seto for providing useful information.
S. K. would like to thank S. Yokoyama for helpful discussions.
T.C. would like to thank T. Asaka and K. Kohri for useful
communications.  This work was supported in part by Grant-in-Aid for
Scientific Research from JSPS [No.20540280(TC), Nos.18072004 and
22340056 (NS)] and in part by Nihon University and World Premier
International Research Center Initiative, MEXT, Japan.


\end{document}